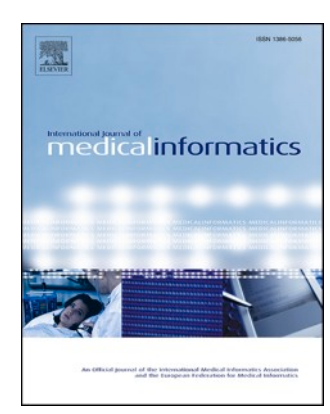

# Logistic regression models for patient-level prediction based on massive observational data: Do we need all data?

Luis H. John [a,*], Jan A. Kors [a], Jenna M. Reps [b], Patrick B. Ryan [b], Peter R. Rijnbeek [a]

[a] *Department of Medical Informatics, Erasmus University Medical Center, Rotterdam, the Netherlands*
[b] *Janssen Research and Development, Raritan, NJ, United States*

ARTICLE INFO

*Keywords:*
Prediction model
Learning curve
Observational data
Sample size
Model complexity
Logistic regression

ABSTRACT

*Objective:* Provide guidance on sample size considerations for developing predictive models by empirically establishing the adequate sample size, which balances the competing objectives of improving model performance and reducing model complexity as well as computational requirements.
*Materials and Methods:* We empirically assess the effect of sample size on prediction performance and model complexity by generating learning curves for 81 prediction problems (23 outcomes predicted in a depression cohort, 58 outcomes predicted in a hypertension cohort) in three large observational health databases, requiring training of 17,248 prediction models. The adequate sample size was defined as the sample size for which the performance of a model equalled the maximum model performance minus a small threshold value.
*Results:* The adequate sample size achieves a median reduction of the number of observations of 9.5%, 37.3%, 58.5%, and 78.5% for the thresholds of 0.001, 0.005, 0.01, and 0.02, respectively. The median reduction of the number of predictors in the models was 8.6%, 32.2%, 48.2%, and 68.3% for the thresholds of 0.001, 0.005, 0.01, and 0.02, respectively.
*Discussion:* Based on our results a conservative, yet significant, reduction in sample size and model complexity can be estimated for future prediction work. Though, if a researcher is willing to generate a learning curve a much larger reduction of the model complexity may be possible as suggested by a large outcome-dependent variability.
*Conclusion:* Our results suggest that in most cases only a fraction of the available data was sufficient to produce a model close to the performance of one developed on the full data set, but with a substantially reduced model complexity.

## 1. Introduction

Physicians infer diagnoses, prognoses, and treatment pathways based on the available medical history of their patients and the current clinical guidelines. Clinical prediction models can support this process by providing risk information on disease presence and progression [1,2]. Well-known prognostic prediction models are QRISK and the Framingham Risk Score, which predict the 10-year risk of developing cardiovascular disease [3,4]. Most such models were developed on small sample sizes and using relatively few predictor variables [2].

In recent years, networks of large observational health databases have opened up possibilities to develop clinical prediction models on massive amounts of patient data. For example, IBM MarketScan Research Databases provide de-identified longitudinal patient-level data of more than 245 million individuals since 1995 [5]. The Observational Health Data Science and Informatics (OHDSI) initiative has recognized the potential of such large observational health databases to develop and validate prediction models for many different populations and health outcomes, and has implemented a standardized and scalable prediction framework [6].

However, previous work has also highlighted limitations of prediction methods when utilizing large amounts of observational data, e.g., tree learning and rule learning algorithms run the risk of having more complexity (e.g. include more features) without achieving substantially better discrimination [7]. As a result, these models may become harder to interpret, more difficult to implement in clinical practice, and more susceptible to overfitting. In addition, developing prediction models on such large data sources can put strong demands on computing resources

* Corresponding author at: Department of Medical Informatics, Erasmus University Medical Center, P.O. Box 2040, 3000 CA Rotterdam, The Netherlands.
*E-mail address:* l.john@erasmusmc.nl (L.H. John).

**Table 1**
Summary of data set characteristics per target cohort and database.

| Target cohort | Database | Target cohort size | No. of events | Outcome rate (%) | No. of candidate predictors |
|---|---|---|---|---|---|
| Depression | MDCD | 654,708* | 8,367 (2,606, 17,223)† | 1.3 (0.4, 2.7)† | 74,469 |
| | MDCR | 371,214 | 7,149 (2,518, 16,697) | 1.9 (0.7, 4.9) | 72,582 |
| | CCAE | 2,893,975 | 17,062 (7,358, 40,797) | 0.6 (0.3, 1.5) | 95,204 |
| Hypertension | MDCD | 144,033 | 2,666 (1,376, 7,157) | 1.8 (0.9, 5.8) | 63,470 |
| | MDCR | 242,298 | 4,382 (1,351, 6,955) | 1.8 (0.5, 3.1) | 60,180 |
| | CCAE | 1,578,418 | 10,228 (4,481, 28,454) | 0.6 (0.3, 1.8) | 89,129 |

*Median; †Median (IQR).

and may require computation times that can become prohibitive.

This work presents an empirical analysis for the development of logistic regression models on massive observational datasets. In particular, we investigate the possibility of reducing the sample size of a large and unwieldy dataset to an "adequate" sample size that is still sufficient to achieve nearly the same performance as the full dataset, which may facilitate the development of less complex clinical prediction models.

An often used heuristic to establish an adequate sample size is to require a minimum number of events per variable (EPV). However, EPV rules have been heavily criticized [8–15]. Moreover, they are not applicable to data-driven modeling approaches in which the number of variables is unknown a priori.

Alternatively, the adequate sample size can be established through a learning curve, which shows model performance as a function of sample size. The sample size at which the plateau phase of the learning curve begins, which indicates convergence towards a maximum performance, can be considered the adequate sample size. To empirically determine the adequate sample size, we have performed one of the largest real-world data studies of predictive modeling to date, fitting 17,248 models through 81 prediction problems (23 outcomes predicted in a depression cohort, 58 outcomes predicted in a hypertension cohort) in three observerational health databases.

### 1.1. Objective

Provide guidance on sample size considerations for developing logistic regression models on massive observational datasets by empirically establishing the adequate sample size, which balances the competing objectives of improving model performance and reducing model complexity as well as computational requirements.

## 2. Materials and methods

### 2.1. Data sources

We used three longitudinal IBM MarketScan Research Databases [5]:

- Commercial Claims and Encounters Database (CCAE) contains data from more than 90 million individuals in the U.S., including employees, their spouses, and dependents enrolled in employer-sponsored insurance health plans between 0 and 65 years of age.
- Multi-State Medicaid Database (MDCD) contains adjudicated U.S. health insurance claims for 44 million Medicaid enrollees from multiple states mostly under 65 years of age.
- Medicare Supplemental Database (MDCR) represents the health services of retirees in the U.S. with primary or Medicare supplemental coverage through privately insured fee-for-service, point-of-service, or capitated health plans. The MDCR population includes 46 million older adults (above 65) and 9 million younger adults with permanent disabilities.

Databases were converted to a standardized data format, the Observational Medical Outcomes Partnership (OMOP) Common Data Model (CDM) [16].

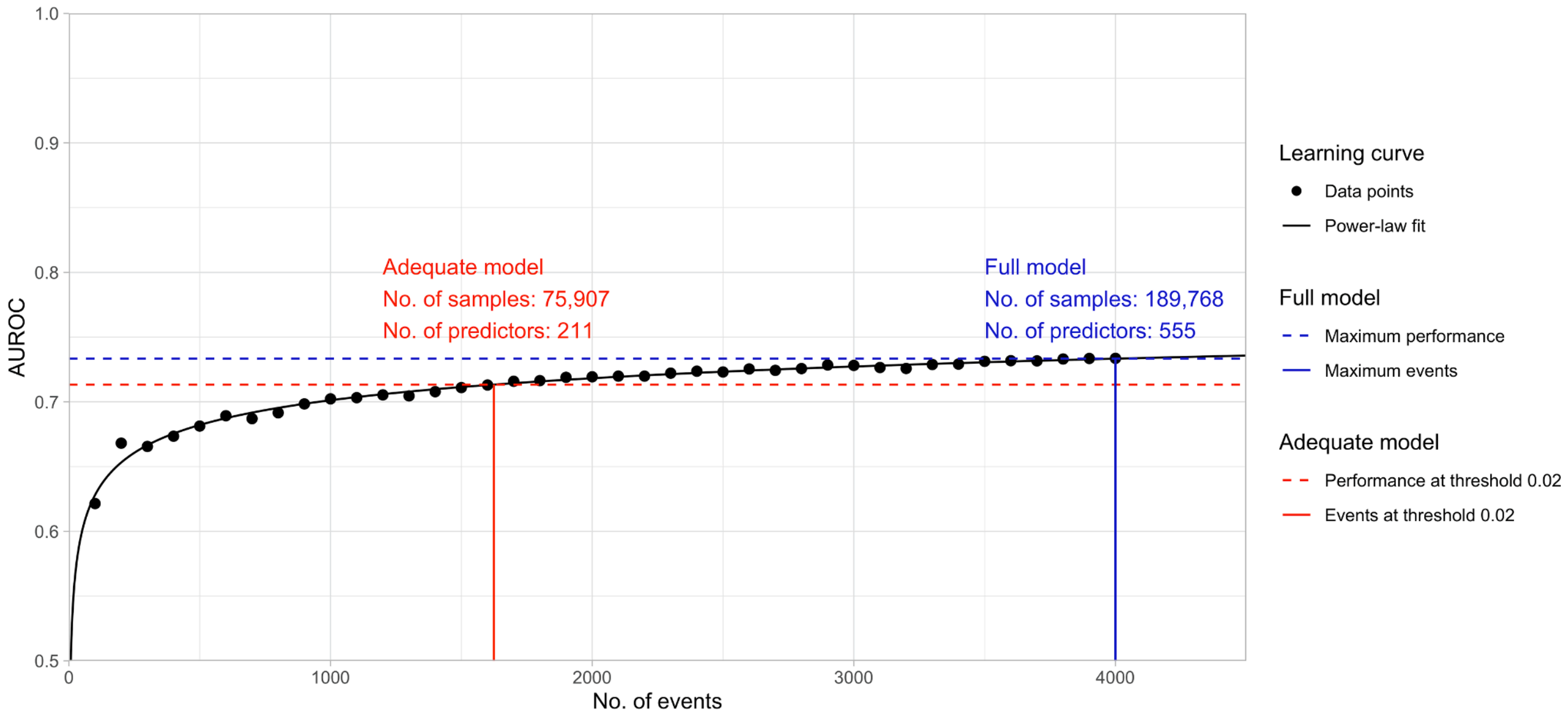

**Fig. 1.** Learning curve for the prediction of venous thromboembolic events in patients with hypertension using data from MDCR. The horizontal lines indicate the maximum performance of the fitted curve (blue) and the performance at a threshold of 0.02 (red). The vertical lines denote the maximum number of events (blue) and the adequate number of events (red). Number of samples and predictors shown for the adequate model pertain to the model at the closest data point. (For interpretation of the references to colour in this figure legend, the reader is referred to the web version of this article.)

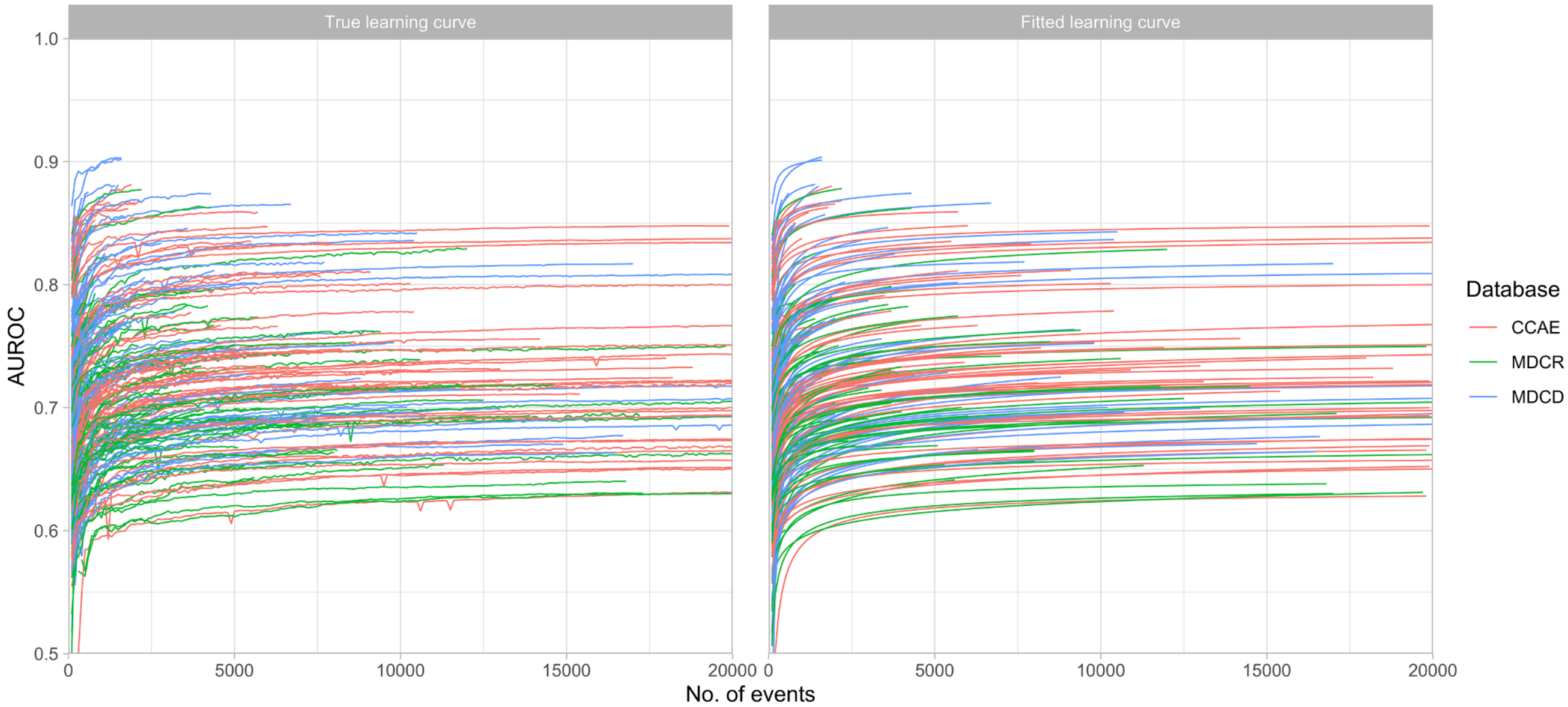

**Fig. 2.** Comparison of the actual learning curves (labeled as true learning curve) on the left and learning curves fitted with the power-law on the right.

**Table 2**
Residual goodness of fit of all included learning curves. The residual is computed as the difference between the true learning curve and the fitted learning curve. The residual is measured in AUROC.

| No. of events | No. of learning curves | Residual |
|---|---|---|
| 1,000 | 200 | 0.0006 (−0.0014, 0.0019)* |
| 2,000 | 171 | 0.0003 (−0.0010, 0.0015) |
| 5,000 | 116 | 0.0001 (−0.0006, 0.0009) |
| 10,000 | 70 | −0.0002 (−0.0006, 0.0002) |
| 20,000 | 23 | 0.0001 (−0.000, 0.0004) |

* Median (interquartile range).

### 2.2. Prediction

We define patient-level prediction as a modeling process wherein a health outcome is predicted within a time-at-risk period relative to the index date of a target cohort. Predictors are derived from patient data in an observation window before the index date. Target cohorts are defined by an index rule specifying the index date, accompanied by a set of inclusion criteria. Criteria can reference one or more clinical concepts that arise from standardized vocabularies in the OMOP CDM [6].

For our analysis, we used 81 previously-defined prediction problems: 23 health outcomes predicted in a depression cohort [6], and 58 health outcomes predicted in a hypertension cohort [17]. Target cohort definitions are detailed in Appendix A. Among the various health outcomes that are predicted in a time-at-risk of 365 days after the index date are, for example, gastrointestinal bleeding, stroke, dementia, and chronic kidney disease. Appendix B provides a list of all prediction problems and target cohort sizes.

#### 2.2.1. Candidate predictors

Candidate predictors include demographics (gender, five-year age groups, race, ethnicity, and index month), condition diagnoses, drug exposures, procedures performed, measurement occurrences, observations, and device exposures.

For all candidate predictors, we determined their occurrence in a long-term window of 365 days before the index date. All non-demographic predictors were additionally considered in a short-term window of 30 days before the index date to distinguish long-term and short-term trends predictive of an outcome.

Table 1 shows the number of candidate predictors and the median target cohort size, number of outcome events, and outcome rate, for each target cohort and database. Target cohort size could differ per health outcomes as detailed in Appendix A.

The five-year age groups were derived by binning and one-hot encoding numerical age to allow modeling of non-linear effects.

#### 2.2.2. Statistical analysis method

For this analysis, we used logistic regression with L1 regularization, also referred to as the least absolute shrinkage and selection operator (LASSO) [18]. LASSO is a regularization method that performs data-driven variable selection where model complexity (taken as the number of predictors) is considered in addition to model performance, generally resulting in a simpler model with fewer predictors. We used the R package Cyclops to perform large-scale regularized logistic regression with cyclic coordinate descent [19]. Importantly, Cyclops performs an automatic search to find the best LASSO regularization parameter during cross validation, eliminating the need for a researcher to manually optimize this parameter. The model's discrimination performance was assessed using the area under receiver operating characteristic curve (AUROC).

### 2.3. Learning curve

To generate a learning curve, we trained prediction models on successively larger subsets of a training set and measured their performance on a test set.

The test set consisted of a random selection of 20% of the full data set. We used the same test set to assess the performance of all models of a learning curve. The remaining 80% of the data was used as the training set and a sequence of proper subsets was defined, where each larger subset contained all data from the smaller subsets.

We report our subset sizes in terms of the number of events they contain analogous to EPV rules, which regard the number of events as the determinant for model performance. We confirmed the strong correlation of the number of events with the learning rate of learning curves in an experiment described in Appendix C.

Our sampling strategy takes each subset by stratified sampling to maintain the same outcome rate as in the complete training set. The size of the subsets increases in steps of 100, from 100 events to 20,000 events (if available).

We evaluate the model's discrimination performance with the area under the receiver operating characteristic curve (AUROC). To draw the learning curves, we plotted AUROC on the test set against the number of events used for training.

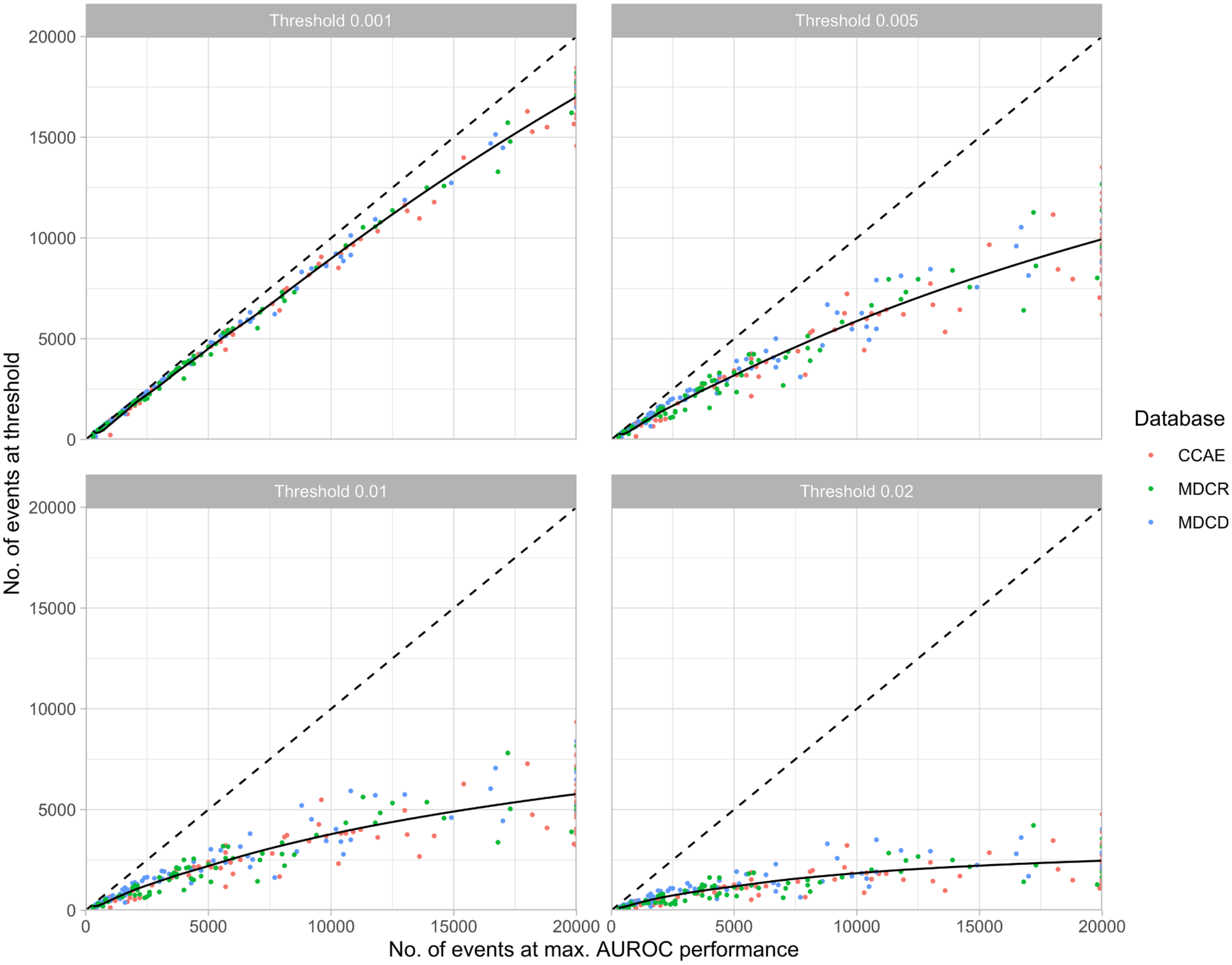

**Fig. 3.** Number of events at thresholds of 0.001, 0.005, 0.01, and 0.02 plotted against the number of events at maximum AUROC performance for the 227 included learning curves.

### 2.3.1. Model complexity curve

In our learning curve analysis we also record the number of predictors selected by LASSO for each model. To draw the model complexity curve, we plotted the number of predictors against the number of events used for training.

### 2.3.2. Power-law fit

Learning curves exhibit some degree of noise due to random effects in the data and model fitting process. To reliably apply convergence criteria, comparative studies have shown that learning curves are best fit with an inverse power law function, which presents an alternative to bootstrapping or resampling techniques when dealing with noise [20–23]. The performance $P$ as a function of the number of events $x$ is then given by [24]:

$$P(x; a, b, c) = (1 - a) - b{\cdot}x^{-c} \tag{1}$$

in which the parameters $a$, $b$, and $c$ represent the minimum achievable error, learning rate, and decay rate, respectively. The fitted curve provides a smoothed, strictly monotone version of the original learning curve.

For fitting the learning curves with the inverse power-law function, we used the Levenberg-Marquardt algorithm, which requires at least as many data points as the number of parameters to fit. Therefore, to fit the three parameters $a$, $b$, and $c$ of $P(x; a, b, c)$, we need at least three data points. Considering a learning curve step size of 100 events between data points, we required at least 300 events in the training set to fit a learning curve with the inverse power-law function.

## 2.4. Determination of the adequate number of events and sample size

To detect learning-curve convergence, slope-based and threshold-based criteria have been proposed. Slope-based criteria determine the slope of the learning curve for different sample sizes and take the smallest sample size for which the slope of the learning curve is "close to zero" [24–28]. However, none of the studies that proposed slope-based criteria quantify when the slope should be considered sufficiently close to zero.

Threshold-based criteria are based on the performance difference between a model trained on all available samples and a model trained on a reduced sample size. The minimum sample size for which this difference becomes smaller than a user-specified threshold is considered adequate [29]. Previous studies that used accuracy as the performance metric considered threshold values varying from 0.001 to 0.02 [7,20,29].

Our approach to determine the adequate number of events relies on the threshold-based method and is exemplified in Fig. 1. for a single prediction problem. In this example, 40 prediction models were developed with number of events between 100 and 4,000. The performances of the models were fitted with a power law function. The performance according to the fitted curve at the maximum number of events was taken as the maximum achievable performance. The number of events where the performance decreased by a prespecified threshold value was considered the adequate number of events. The corresponding adequate sample size was computed by dividing by the outcome rate of the prediction problem. We repeated this exercise for 81 target-outcome pairs, within 3 different datasbases.

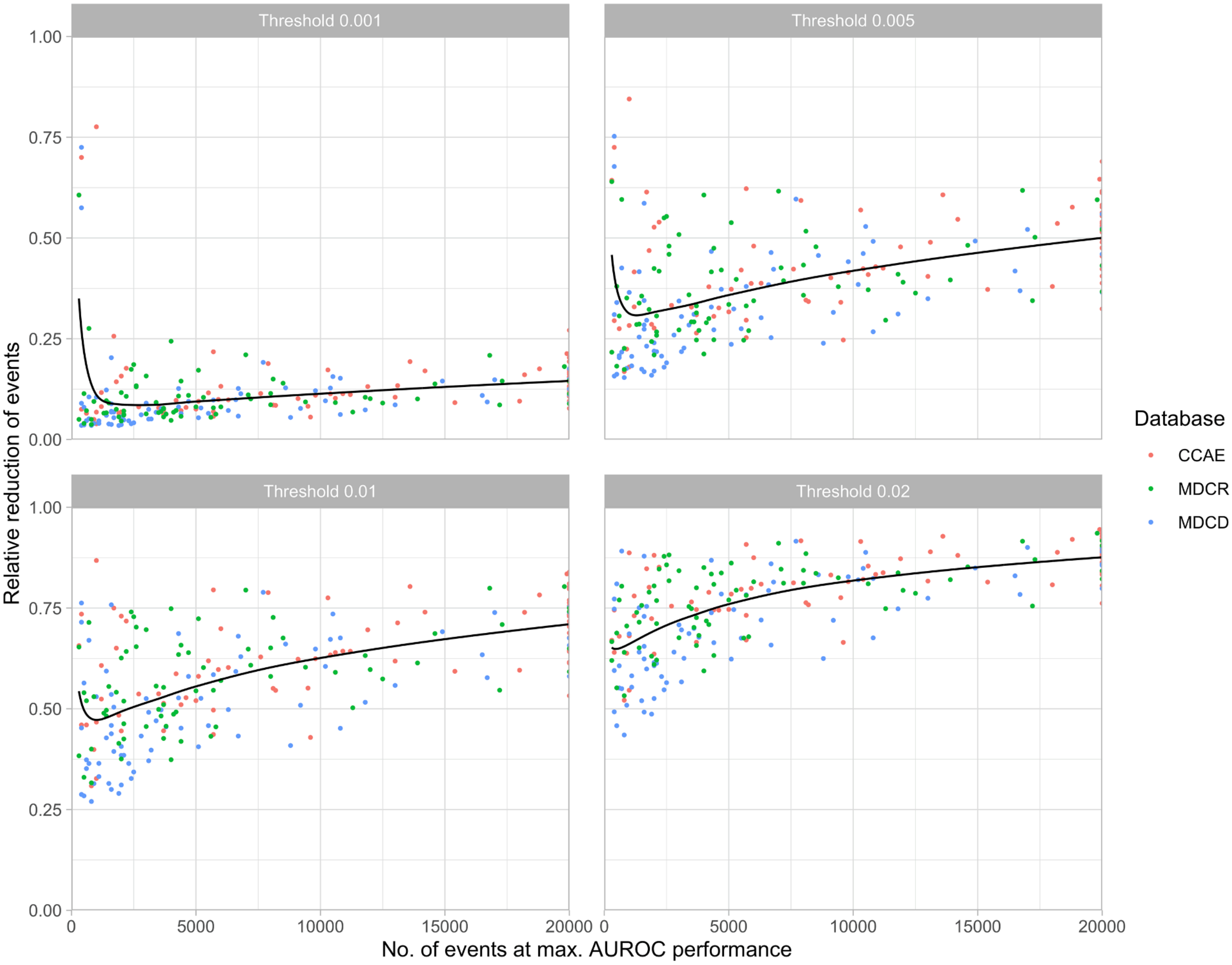

**Fig. 4.** Relative reduction of events at thresholds of 0.001, 0.005, 0.01, and 0.02 plotted against the number of events at maximum AUROC performance for the 227 included learning curves.

We considered threshold values of 0.001, 0.005, 0.01, and 0.02. For each adequate number of events (Na), we computed the reduction of the number of events relative to the maximum number of events (N): $100 \cdot \frac{(N-Na)}{N}$.

## 3. Results

From the 81 prediction problems, we excluded those that had fewer than 300 events in the training set, resulting in 80, 75, and 72 prediction problems for CCAE, MDCD, and MDCR, respectively. Using these datasets, a total of 17,248 prediction models were trained and evaluated to generate the learning curves for the 227 included prediction problems. Fig. 2 visualizes the true and fitted learning curves. While many of the learning curves appear to be plateauing, some have not yet reached that phase. The median (interquartile range) of the maximum AUROCs of the prediction problems was 0.742 (0.700–0.790). As may be expected, the instability of the true learning curves is highest for small number of events and gradually diminishes with increasing number of events. The individual learning curves can be found in Appendix D.

To evaluate the goodness of fit, we computed the residuals as the AUROC performance difference between the true learning curve and the fitted learning curve (Table 2). The median and interquartile range of the residuals show that the fit is excellent. It is noteworthy that increasingly fewer data sets could provide larger number of events apparent from the number of included learning curves. Appendix E provides additional goodness of fit results for included learning curves.

### 3.1. *Adequate sample size*

Fig. 3 shows the adequate number of events of the prediction problems against the number of events at maximum performance for the various thresholds. A substantial reduction of the number of events can be achieved, although there exists considerable variation around the trend lines, indicating different degrees of reduction for different prediction problems and databases.

The relative reduction of the number of events at the four thresholds is shown in Fig. 4. Of note, since this reduction is a proportion, the relative reductions in number of events and in full sample size are the same. There is an upward trend for larger number of events, which indicates that the reduction is more substantial for larger data sets. For the thresholds of 0.001, 0.005, 0.01, and 0.02 we achieve a median reduction (interquartile range) of 9.5% (6.7%–13.6%), 37.3% (28.4%–48.0%), 58.5% (46.1%–68.8%), and 78.5% (67.8%–85.1%) respectively. There is a large variability of the relative reduction across the different prediction problems and databases indicated by the vertical spread of the data points.

### 3.2. *Model complexity*

We assessed how model complexity changes at the threshold-based adequate number of events as compared to the full models based on all available of events. Because at the adequate number of events no datapoint was generally recorded (model complexity curve was not fitted), we decided to report the complexity of the model trained on the next smallest subset as the adequate model. Prediction problems were omitted from this plot, if the adequate model had less than 100 events.

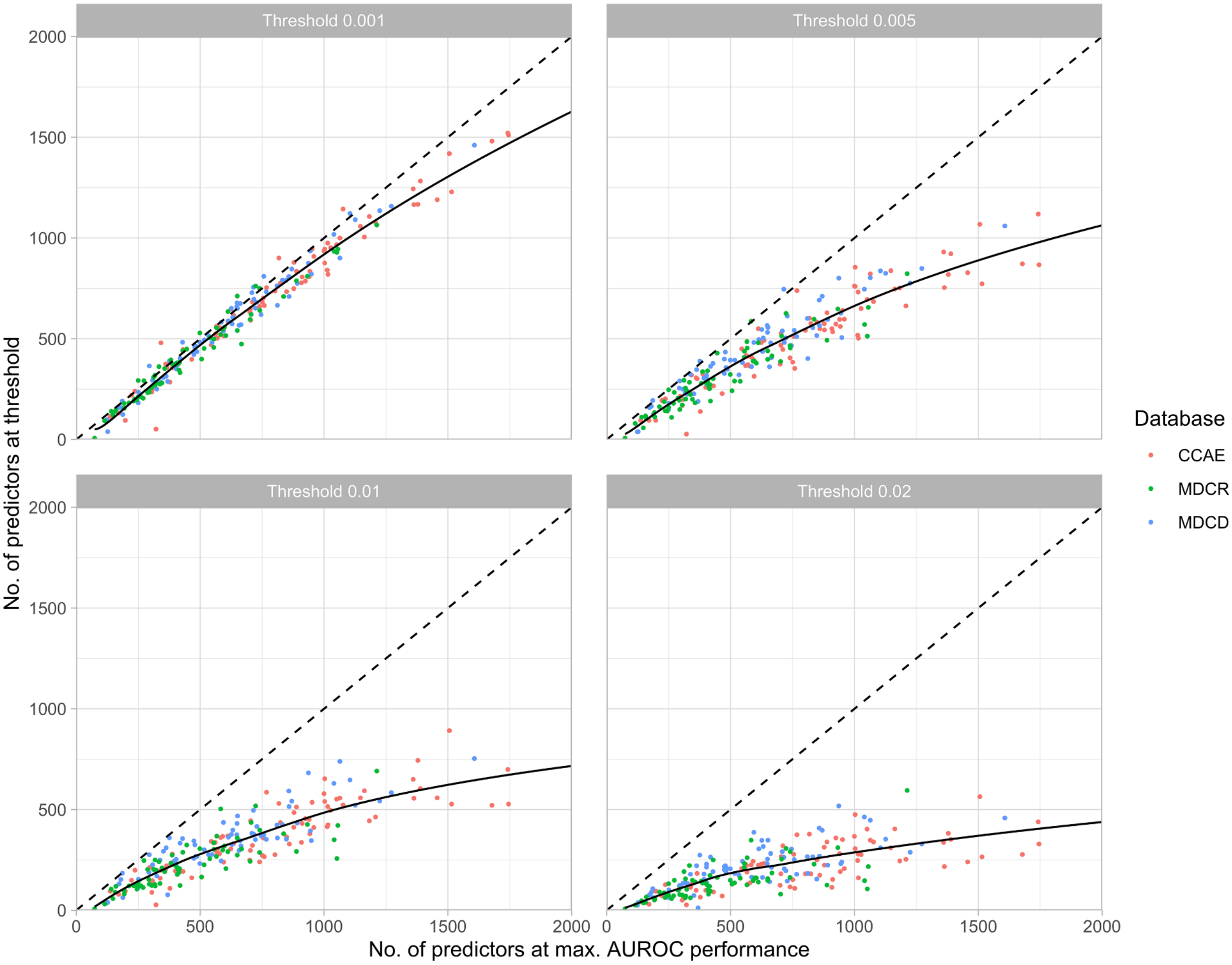

**Fig. 5.** Number of predictors at thresholds of 0.001, 0.005, 0.01, and 0.02 plotted against the number of predictors at maximum AUROC performance for the 227 included learning curves.

Fig. 5 shows for each prediction problem the number of predictors in the full model (based on the maximum number of events) versus the number of predictors in the adequate model (based on the adequate number of events) for the different threshold values. Full models had a median (interquartile range) of 588 (356–858) predictors, while models at the largest threshold of 0.02 had 181 (120–265) predictors.

This substantial reduction of model complexity is achieved in addition to the sample size reduction. Relatively, for the thresholds 0.001, 0.005, 0.01, 0.02 we achieve a median reduction (interquartile range) of model complexity of 8.6% (3.8%–14.6%), 32.2% (22.6%–38.4%), 48.2% (38.7%–56.7%), 68.3% (58.9%–75.7%), respectively (Fig. 6). Similar to the reduction in the number of events, there is a substantial spread indicated by the interquartile ranges.

Appendix F details results of complexity growth in logistic regression models. Additionally, we evaluated and discussed EPV values of our models in Appendix G.

## 4. Discussion

Our extensive analysis of learning curves on large observational health datasets showed that a considerable reduction in sample size and model complexity paired with a minimal loss of performance is possible.

Using a threshold-based convergence criterion we defined the adequate sample size, which achieves a median reduction of the number of events between 9.5% and 78.5% for thresholds between 0.001 and 0.02. Because the outcome rate in the learning curve's subsets was held constant, this reduction in the number of events can be translated to a reduction in the absolute sample size of the same proportion. Remarkably, also for small datasets often large reductions are feasible. However, in general larger datasets saw larger reductions as indicated by the trend lines in Fig. 4.

An additional benefit of using the adequate sample size is a less complex model. We could achieve a median reduction of the number of predictors between 8.6% and 68.3% for the thresholds between 0.001 and 0.02. Oates et al. have previously reported that larger datasets lead to overly complex decision tree models [7]. Our experiments show that this phenomenon also holds true for logistic regression models fitted using LASSO. Therefore, using the adequate sample size we can develop less complex logistic regression models, which has implications for applying these models in practice. For example, a large number of predictors can be a barrier to clinical implementation when a database cannot provide such features. Fewer predictors may also become more interpretable for clinicians and easier to apply manually. Future research should be dedicated to assessing the adequate model's transportability compared to a full model through external validation.

In light of our results using logistic regression, other prediction methods utilizing deep learning will also immediately come to mind when massive data sets are implied, since recent accomplishments in the fields of medical image analysis, computational genomics, but also disease prediction are numerous [30–32]. However, deep learning is not a universal remedy for all healthcare analytics problems [33].

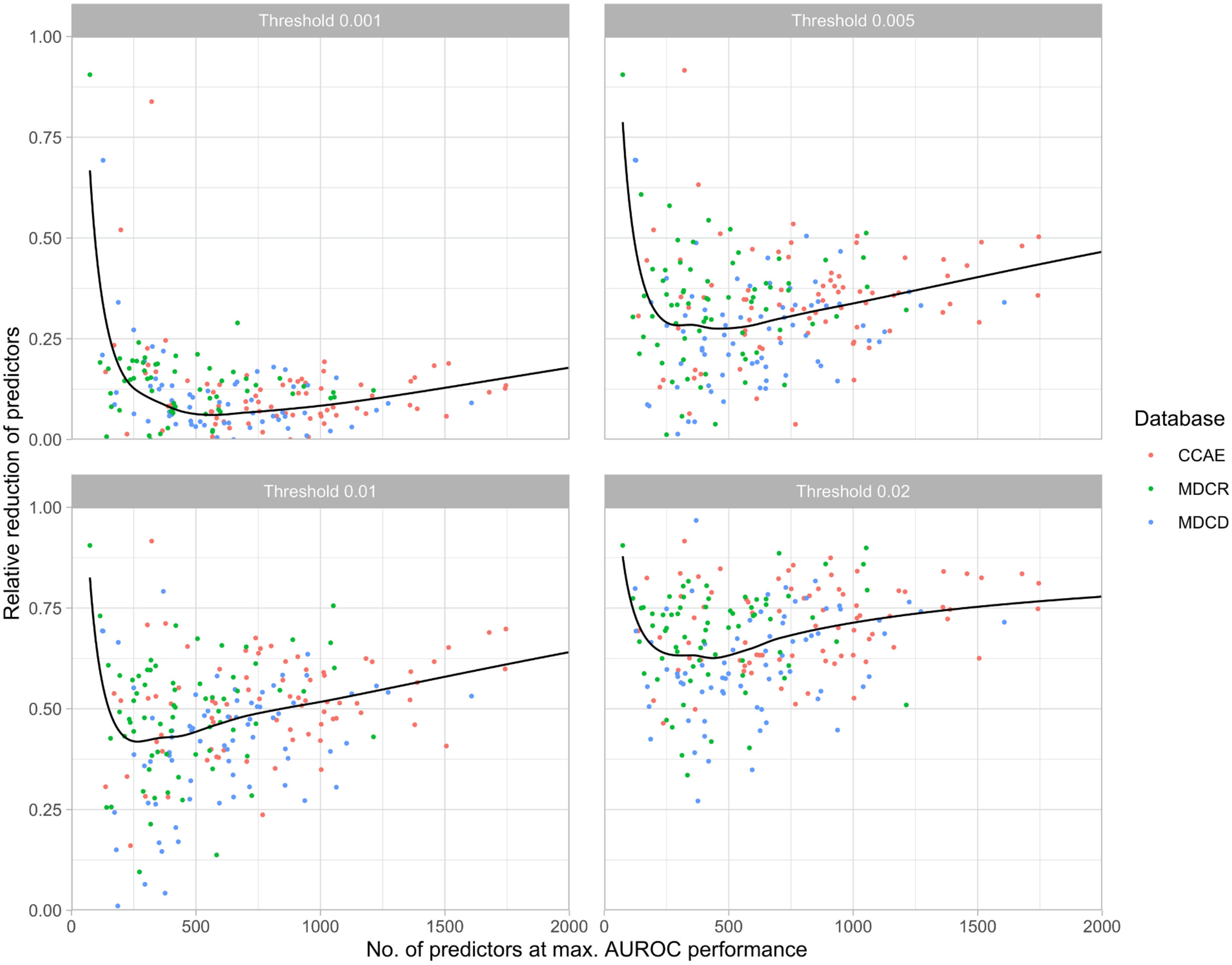

**Fig. 6.** Relative reduction of predictors at thresholds of 0.001, 0.005, 0.01, and 0.02 plotted against the number of predictors at maximum AUROC performance.

Observational patient-level data exhibits properties such as high sparsity, high dimensionality, and heterogeneity, that have been found to limit the efficacy of deep learning methods [33–36]. Conventional machine learning methods could be trained faster and had overall better performance when compared to deep neural networks [33]. Despite these potential obstacles, we believe that an empirical assessments of learning curves generated from deep learning models may present an exciting opportunity for future research.

### *4.1. Use of learning curves*

Comparative studies have shown that learning curves are best fit with a power-law [20–23]. We also found the power-law fit to be excellent, with median residuals smaller than 0.001. However, the interquartile range of the fit is in the order of our smallest threshold value of 0.001, indicating that our results for this threshold may need to be interpreted with caution.

Future work is required to deepen our understanding of learning curves, for example, by comparing how the adequate sample size behaves for different metrics (e.g. AUPRC) and different prediction algorithms (e.g. random forest).

Another research question is 'how much performance can be gained?', if more data were available. This is an extrapolation question for which fitted learning curves may also be useful.

### *4.2. Recommendations*

The possibility of reducing sample size and model complexity can have implications for prediction studies. Depending on their goals, researchers may decide for one of two scenarios:

(1) If the goal is to reduce computational resources with the additional benefit of reduced model complexity, we recommend sample size reduction based on the minimum expected relative reduction in the number of events as shown in Fig. 4. For example, if a researcher accepts a threshold of 0.01, the median reduction for a given data set with 20,000 events would be 73%, and thus sample size could be reduced with 73%, corresponding to a median complexity reduction of 59% (Fig. 6). For this approach no learning curve needs to be generated, which saves on the overhead associated with fitting the various models.
(2) If the goal is to reduce model complexity for a given data set as much as possible, we recommend a learning curve is generated to determine the adequate sample size. Generating a learning curve will increase the use of computational resources but the complexity reduction is potentially much larger than in the previous scenario. For example, for a threshold of 0.01 and a data set with 20,000 events, Fig. 4 shows that a sample size reductions of up to 84% may be possible.

### *4.3. Limitations*

Prediction problems in this study originated from only two target cohorts (depression and hypertension), in which the various health outcomes were predicted, and three databases, which exclusively held U.S. administrative claims data. Moreover, only a single prediction algorithm was assessed, which was logistic regression. As a result there may be limitations for scenario (1) in terms of generalizability to other target populations, data sources and prediction algorithms. More problem specific sample size recommendations can be received through generating a learning curve as proposed in scenario (2).

## 5. Conclusion

In this study we empirically derived the adequate sample size for logistic regression models using learning curves. Our results suggest that in most cases only a fraction of the available data was sufficient to produce a model close to the performance of one developed on the full data set, but with a substantially reduced model complexity.

## Contributorship statement

L.H.J., J.A.K. and P.R.R. conceived and planned the experiments. L.H.J. prepared the data and carried out the experiments. L.H.J., J.A.K., P.B.R., J.M.R. and P.R.R. contributed to the interpretation of the results. L.H.J. took the lead in writing the manuscript. All authors provided critical feedback and helped shape the research, analysis and manuscript.

## Funding statement

This work has received funding from the Innovative Medicines Initiative 2 Joint Undertaking (JU) under grant agreement No. 806968. The JU receives support from the European Union's Horizon 2020 research and innovation programme and EFPIA.

## Summary Table

| What was already known on the topic? | What this study added to our knowledge |
|---|---|
| • Massive observational health datasets have become available, enabling large-scale development and validation of prediction models<br>Data-driven approaches to predictive modeling lack guidelines to estimate adequate sample sizes beyond which model performance does not increase | • For regularized logistic regression models developed on large observational health datasets, the adequate sample size is often considerably smaller than the full sample size<br>Complexity of logistic regression models can be decreased by reducing the sample size of large observational datasets |

## Declaration of Competing Interest

The authors declare that they have no known competing financial interests or personal relationships that could have appeared to influence the work reported in this paper.

## Appendix A. Supplementary data

Supplementary data to this article can be found online at https://doi.org/10.1016/j.ijmedinf.2022.104762.